\documentstyle[12pt]{article}
\newcommand{\bfk}{{\mbox{\boldmath $k$}}}
\newcommand{\bfp}{{\mbox{\boldmath $p$}}}
\newcommand{\bfB}{{\mbox{\boldmath $B$}}}
\newcommand{\bfsig}{{\mbox{\boldmath $\sigma$}}}
\newcommand{\bfn}{{\mbox{\boldmath $n$}}}

\newcommand{\bfq}{{\mbox{\boldmath $q$}}}
\newcommand{\one}{{\mbox{1\hskip-0.5mm l}}}
\begin{document}
\begin{center}
{\large \bf
Spin Physics with Top Quarks at Hadron Colliders
\\ }
\vspace{5mm}
Arnd Brandenburg
\\
\vspace{5mm}
{\small\it
Institut f\"ur Theoretische Physik,
RWTH Aachen,
52056 Aachen, Germany 
\\}
\end{center}

\begin{center}
ABSTRACT

\vspace{5mm}
\begin{minipage}{130 mm}
\small
We discuss the prospects to observe effects 
of transverse polarization and spin-spin correlations 
of top quark pairs produced at hadron colliders. 
\end{minipage}
\end{center}
The detailed study of the properties of the top quark will form 
an important part of the experimental program at the (upgraded) 
Tevatron and at future colliders.  
A special feature of the top quark is that due to its large mass 
it decays on average before it can hadronize. Moreover, the typical 
spin flip time is much larger than the lifetime of the top. 
Thus a $t$($\bar{t}$) polarization and spin-spin correlations 
between the $t$ and $\bar{t}$ induced by the
production mechanism will be transferred to the decay products.
The spin information may be extracted by forming angular correlations 
among the $t$ and $\bar{t}$ decay products, 
thus allowing for a variety of tests of the standard
model and its extensions (see, e.g. [1-5] and references therein).
\par
It is  useful to introduce the concept of a {\it{spin density matrix}} 
for the $t\bar{t}$ system. 
As an example we discuss the reaction $q(p_1)+\bar{q}(p_2)\to
t(k_1)+\bar{t}(k_2)$, where the momenta refer to the partonic c.m.
system. This reaction is the dominant production
mechanism for top quark pairs at Tevatron energies.
  The complete spin information is encoded
in the (unnormalized) spin density matrix $R^q$,
\begin{eqnarray} \label{density1}
R^q_{\alpha_1\alpha_2,\beta_1\beta_2}(\bfp,\bfk)=
\frac{1}{4}\frac{1}{N_C^2}\sum_{{\mbox{\scriptsize{colors}}},
q\bar{q}{\mbox{\scriptsize{\ spins}}}}
\langle t(k_1,\alpha_2)\bar{t}(k_2,\beta_2)|{\cal{T}}|
q(p_1),\bar{q}(p_2)\rangle^* & &\nonumber\\
\langle t(k_1,\alpha_1)
\bar{t}(k_2,\beta_1)|{\cal{T}}|
q(p_1),\bar{q}(p_2)\rangle. & &
\end{eqnarray}
Here, $\alpha,\ \beta$ are spin indices, $N_C$ denotes the number of
colors,  $\bfp=\bfp_1,\ \bfk=\bfk_1$, and the sum runs over the colors
of all quarks and over the spins of $q$ and $\bar{q}$. The factor
$1/4\cdot 1/N_C^2$ takes care of the averaging over spins and colors
in the initial state. The matrix structure of $R^q$ in the spin spaces
of $t$ and $\bar{t}$ is
\begin{equation}
R^q=A^q\one\otimes \one+ \bfB^q_t\cdot\bfsig\otimes \one+\bfB^q_{\bar{t}}\cdot
\one\otimes\bfsig+C^q_{ij}\sigma^i\otimes\sigma^j,
\end{equation}
where $\sigma^i$ are the Pauli matrices and the first (second) factor
in the tensor products refers to the $t$ ($\bar{t}$) spin space. 
A more detailed discussion of $R^q$ is given in [1]. The Born result for
this matrix and the corresponding one $R^g$ for the reaction 
$gg\to t\bar{t}$ can be found in a compact form in [2]. 
\par
\vspace{0.2cm}
\noindent{\bf Transverse Polarization}
\par
While at hadron colliders the 
{\it{longitudinal}} polarization of top quarks is practically zero due
to parity invariance of QCD, a nonzero polarization {\it{transverse}}
to the production plane is induced by absorptive parts 
in the parton scattering amplitude of
$q\bar{q}\to t\bar{t}$ and $gg\to t\bar{t}$ [3,4]. 
They yield the following contribution to the production density
matrices $R^{q,g}$: 
\begin{equation}
R_{{\mbox{\scriptsize{abs}}}}^{q,g}=b_3^{q,g}(\hat{s},z)\hat\bfn\cdot
\left\{\bfsig\otimes \one+\one\otimes\bfsig\right\},
\end{equation}
where $\hat{s}=(p_1+p_2)^2$, $z=\hat\bfp\cdot\hat\bfk$ is the cosine
of the scattering 
angle in the partonic center of mass frame, 
$\hat\bfn=(\bfp\times\bfk)/|\bfp\times\bfk|$ is the normal to the
scattering plane, and $b_3^{q,g}(\hat{s},z)$ is a structure function in
the notation of [1]. 
The transverse polarization  $P_{\bot}$ 
is equal for $t$ and $\bar{t}$ 
and given for the respective subprocesses by
\begin{equation}
P_{\bot}^{i}=\langle \hat\bfn\cdot\bfsig\otimes\one\rangle_{i}
\equiv\frac{{\mbox{tr}}(R^{i}
\ \hat\bfn\cdot\bfsig\otimes\one)}{{\mbox{tr}}
\ R^{i}}=\frac{b_3^{i}(\hat{s},z)}{A^{i}(\hat{s},z)}\ \ \ \ \ \ (i=q,g).
\end{equation}
The functions $A^{i}(\hat{s},z)$ are related to the differential 
cross section
of the respective partonic subprocess.
For quark--antiquark 
annihilation, the transverse polarization of the top quark 
reaches values of about $2.5\%$ around $\sqrt{\hat{s}}\simeq 720$ GeV 
and scattering angle of $\simeq 73^{\circ}$   and then 
decreases quite
rapidly with energy. In the case of gluon--gluon fusion, $P_{\bot}^g$ 
reaches peak values of about $1.5\%$  
around $\sqrt{\hat{s}}\simeq 1050$ GeV and at $\simeq \pm 63^{\circ}$. 
(For the above numbers we put $m_t=180$ GeV and $\alpha_s=0.1$.)  
The transverse polarization of $t$ and $\bar{t}$ must be traced 
in the final states into which $t$ and 
$\bar{t}$ decay. Especially suited are ``semihadronic'' decay modes, i.e.
decay modes  where one of the
top quarks decays semileptonically and the other one decays hadronically,
$t\to \ell^+\nu_{\ell}b$, 
$\bar{t} \to q\bar{q}'\bar{b}$
and the corresponding charge conjugated decay channels. 
For $p\bar{p}\to t\bar{t}X$
with subsequent semileptonic
$t\bar{t}$ decay we define the observable
\begin{equation}
{\cal O}_1= \hat\bfp_p\cdot(\hat{\bfq}_{{\ell}^+}
\times \hat{\bfk}_{\bar{t}}) 
\end{equation}
for the decay modes given above and
\begin{equation}
\bar{\cal O}_1= \hat\bfp_p\cdot(
\hat{\bfq}_{{\ell}^-}\times \hat{\bfk}_{t}) \label{Om}
\end{equation}
for the charge conjugated decays. Here, $\hat\bfp_p$ is the 
proton's direction, $\bfq_{\ell^{\pm}}$ is the momentum
of the positively (negatively) charged lepton,
hats denote unit vectors  and all momenta
are defined in the hadronic c.m. system. 
A correlation sensitive to transverse polarization may now be defined 
through the sum:
\begin{equation}
{\cal C}_1=\langle{\cal O}_1\rangle_{\bar{t}\ell^{+}}+
         \langle\bar{\cal O}_1\rangle_{t\ell^{-}}.
\end{equation}  
At $\sqrt{s}=1.8$ TeV the correlation has the value
${\cal C}_1\simeq 0.43$\% and it decreases  with rising energy.
The statistical sensitivity of the observables may be estimated
from their root-mean-square fluctuations; 
we find for the  1 s.d. statistical
error $\delta {\cal C}_1\simeq 0.5/\sqrt{N_{b\ell^-}}$,
where $N_{b\ell^-}$ is the number of events of the ``semihadronic'' type.
In order to push $\delta {\cal C}_1$ below the
percent level $N_{b\ell^-}> 10^4$ events 
are needed  -- an unrealistic 
number even for the upgraded
Tevatron. For $pp$ collisions at LHC energies the effects are even smaller,
making the correlation (7) potentially sensitive to new physics effects.
\par
\vspace{0.2cm}
\noindent{\bf Spin-Spin Correlations}
\par
Apart from the small single quark polarization discussed above, 
the $t$ and $\bar{t}$ are
produced with {\it{strongly correlated spins}}. In
fact, these spin-spin correlations are of order one at the level of
the partonic reactions. They can be described by four structure functions
which determine $C_{ij}^{q,g}$ defined in (2),
\begin{equation}
C_{ij}^{q,g} = c_0^{q,g}(\hat{s},z)\delta_{ij}+ 
c_4^{q,g}(\hat{s},z)\hat{p}_i\hat{p}_j+c_5^{q,g}(\hat{s},z)\hat{k}_i\hat{k}_j
+c_6^{q,g}(\hat{s},z)(\hat{p}_i\hat{k}_j+\hat{k}_i\hat{p}_j).
\end{equation} 
Observables built from the spin
operators  ${\bf s}_{t(\bar{t})}$ of $t$ and $\bar{t}$ 
allow to discuss the magnitude of the
correlations at parton level. For example, at the Born level,
$\langle {\bf s}_t\cdot {\bf s}_{\bar{t}}\rangle_{q\bar{q}\to t\bar{t}}=1/4$.
Again these effects must be traced in the $t\bar{t}$ decay products.
Different strategies have been proposed recently [2,5]. We concentrate here 
on observables that can be measured on an event by event basis, again using the
``semihadronic'' decay modes.
At the Tevatron, the observable with the highest statistical significance we 
found is
\begin{equation}
{\cal {O}}_2=(\hat{\bfq}_b^*\cdot\hat{\bfp}_p)
(\hat{\bfq}_{\ell^-}\cdot\hat{\bfp}_p).
\end{equation}
Here, $\hat{\bfq}_b^*$ is the direction of flight of the bottom quark in the
top quark rest frame. 
An analogous
observable $\bar{\cal{O}}_{2}$ may be defined for the charge 
conjugated decay modes. For the statistical sensitivitiy of the
combined correlations
$\langle{\cal{O}}_2\rangle$ and $\langle\bar{\cal{O}}_2\rangle$  we have
\begin{equation}
S_2\equiv\frac{|\langle{\cal{O}}_2\rangle+\langle\bar{\cal{O}}_2\rangle|}
{\sqrt{2}\Delta{\cal{O}}_2}\sqrt{N_{b\ell^-}}\approx 0.14\sqrt{N_{b\ell^-}}
\end{equation}
at $\sqrt{s}=1.8$ TeV. In order to establish the spin-spin correlation
at the 3$\sigma$ level, we would therefore need $N_{b\ell^-} \approx
500$, which is in reach of the upgraded Tevatron.
For the LHC, the ``best'' observable is given by
\begin{equation}
{\cal{O}}_3 = (\hat{\bfp}_p\times \hat{\bfq}_b^*)\cdot
                  (\hat{\bfp}_p\times \hat{\bfq}_{\ell^-}).
\end{equation}
We find for the combined correlations $\langle{\cal{O}}_3\rangle$ and
 $\langle\bar{\cal{O}}_3\rangle$ a statistical sensitivity
of ${\cal{S}}_3\approx 0.073 \sqrt{N_{b\ell^-}}$ at $\sqrt{s}=14$ TeV. 
 Assuming  $N_{b\ell^-}=10^4=N_{\bar{b}\ell^+}$ , 
we can therefore establish the spin-spin correlations to $7.3\sigma$
by measuring ${\cal{O}}_3$ and $\bar{\cal{O}}_3$.
\par
In summary, interesting spin physics with top quarks is possible 
at the upgraded Tevatron and the LHC. 
\vspace{0.2cm}
{\small\begin{description}
\item{[1]} W. Bernreuther and A. Brandenburg,
  Phys. Lett. {\bf B 314} (1993) 104; Phys. Rev. {\bf D 49} (1994)
  4481. 
\item{[2]}
A. Brandenburg, PITHA 96/07, hep-ph/9603333, to appear in Phys. Lett. {\bf B}.
\item{[3]} 
W. Bernreuther, A. Brandenburg, and P. Uwer,
Phys. Lett. {\bf B 368} (1996) 153.
\item{[4]}
 W. G. D. Dharmaratna and G. R. Goldstein, Phys. Rev. {\bf D 53} (1996) 1073.
\item{[5]} G. Mahlon and S. Parke, Phys. Rev. {\bf D 53} (1996) 4886;
D. Chang, S. Lee, and A. Soumarokov, Phys. Rev. Lett. {\bf 77} (1996) 1218.
\end{description}}

\end{document}